\documentclass[aps,prd,onecolumn,groupedaddress,showpacs,nofootinbib,amssymb]{revtex4}
\usepackage[dvips]{graphicx}
\usepackage{amssymb}
\usepackage{amsmath}
\usepackage{graphicx,,color}
\usepackage{amsfonts}
\usepackage{bm}
\usepackage{cancel}
\usepackage{comment}

\newcommand\be{\begin{equation}}
\newcommand\ee{\end{equation}}

\allowdisplaybreaks[4]

\begin{document}

\tolerance=5000

\title{Rectifying Einstein-Gauss-Bonnet Inflation in View of GW170817}
\author{S.D.~Odintsov,$^{1,2}$\,\thanks{odintsov@ice.cat}
V.K.~Oikonomou,$^{3,4,5}$\,\thanks{v.k.oikonomou1979@gmail.com}F.P.
Fronimos,$^{3}$\,\thanks{fotisfronimos@gmail.com}}
\affiliation{$^{1)}$ ICREA, Passeig Luis Companys, 23, 08010 Barcelona, Spain\\
$^{2)}$ Institute of Space Sciences (IEEC-CSIC) C. Can Magrans
s/n,
08193 Barcelona, Spain\\
$^{3)}$ Department of Physics, Aristotle University of
Thessaloniki, Thessaloniki 54124,
Greece\\
$^{4)}$ Laboratory for Theoretical Cosmology, Tomsk State
University of Control Systems and Radioelectronics, 634050 Tomsk,
Russia (TUSUR)\\
$^{5)}$ Tomsk State Pedagogical University, 634061 Tomsk,
Russia\\}

 \tolerance=5000

\begin{abstract}
In this work we introduce a new theoretical framework for
Einstein-Gauss-Bonnet theories of gravity, which results to
particularly elegant, functionally simple and transparent
gravitational equations of motion, slow-roll indices and the
corresponding observational indices. The main requirement is that
the Einstein-Gauss-Bonnet theory has to be compatible with the
GW170817 event, so the gravitational wave speed $c_T^2$ is
required to be $c_T^2\simeq 1$ in natural units. This assumption
was also made in a previous work of ours, but in this work we
express all the related quantities as functions of the scalar
field. The constraint $c_T^2\simeq 1$ restricts the functional
form of the scalar Gauss-Bonnet coupling function $\xi(\phi)$ and
of the scalar potential $V(\phi)$, which must satisfy a
differential equation. However, by also assuming that the
slow-roll conditions hold true, the resulting equations of motion
and the slow-roll indices acquire particularly simple forms, and
also the relation that yields the $e$-foldings number is
$N=\int_{\phi_i}^{\phi_f}\xi''/\xi'd \phi$, a fact that enables us
to perform particularly simple calculations in order to study the
inflationary phenomenological implications of several models. As
it proves, the models we presented are compatible with the
observational data, and also satisfy all the assumptions made
during the process of extracting the gravitational equations of
motion. More interestingly, we also investigated the
phenomenological implications of an additional condition
$\xi'/\xi''\ll 1$, which is motivated by the slow-roll conditions
that are imposed on the scalar field evolution and on the Hubble
rate. As we shall show, the resulting constraint differential
equation that constrains the functional form of the scalar
Gauss-Bonnet coupling function $\xi(\phi)$ and of the scalar
potential $V(\phi)$, is simpler in this case, and in effect the
whole study becomes somewhat easier. As we also show,
compatibility with the observational data can also be achieved in
this case too, in a much simpler and less constrained way. Our
approach opens a new window in viable Einstein-Gauss-Bonnet
theories of gravity.
\end{abstract}

\pacs{04.50.Kd, 95.36.+x, 98.80.-k, 98.80.Cq,11.25.-w}

\maketitle

\section{Introduction}

The last twenty years had a lot of surprises for theoretical
cosmologists, coming from both cosmological scale data and also
from astrophysical scales events. Particularly, the observation of
the currently accelerating Universe coming from the standard
candles SNe Ia \cite{Riess:1998cb}, has utterly changed our
perception of how the Universe evolves. In addition, the direct
detection of gravitational waves coming from the merging of two
neutron stars in 2017 \cite{GBM:2017lvd}, the GW170817 event as it
is widely known, also affected theoretically cosmology
drastically. This is due to the fact that the gravitational waves
arrived almost simultaneously with the gamma rays emitted from the
merging neutron stars event, and this indicated that the
gravitational wave speed is $c_T^2\simeq 1$, in natural units.
This fact, strongly imposed stringent conditions on modified
gravity theories that may successfully describe nature on such
scales, and actually excluded a large number of theories, see Ref.
\cite{Ezquiaga:2017ekz} for a complete list of the theories that
are excluded from being viable, after the GW170817.

In a previous work \cite{Odintsov:2019clh} we demonstrated that it
is possible to make the Einstein-Gauss-Bonnet theories
\cite{Hwang:2005hb,Nojiri:2006je,Cognola:2006sp,Nojiri:2005vv,Nojiri:2005jg,Satoh:2007gn,Bamba:2014zoa,Yi:2018gse,Guo:2009uk,Guo:2010jr,Jiang:2013gza,Kanti:2015pda,vandeBruck:2017voa,Kanti:1998jd,Kawai:1999pw,Nozari:2017rta,Chakraborty:2018scm,Odintsov:2018zhw,Kawai:1998ab,Yi:2018dhl,vandeBruck:2016xvt,Kleihaus:2019rbg,Bakopoulos:2019tvc,Maeda:2011zn,Bakopoulos:2020dfg,Ai:2020peo,Easther:1996yd,Antoniadis:1993jc,Antoniadis:1990uu,Kanti:1995vq,Kanti:1997br}
compatible with the GW170817 event, and making the gravitational
wave speed to be $c_T^2\simeq 1$. Actually, technically this can
be achieved, since in the Einstein-Gauss-Bonnet theories case, the
gravitational wave speed is equal to $c_T^2=1-\frac{Q_f}{2Q_t}$,
where $Q_f=8(\ddot\xi-H\dot\xi)$. Thus if the scalar coupling
function $\xi(\phi)$ is chosen so that it satisfies the
differential equation $\ddot\xi-H\dot\xi=0$, the parameter $Q_f$
becomes identically equal to zero. The approach we adopted in
\cite{Odintsov:2019clh}, was cosmic time oriented, and the results
were obtained by using in most cases expressions involving the
cosmic time. However, we realized that the GW170817-compatible
Einstein-Gauss-Bonnet inflationary theory might be developed in a
much more simple and transparent way if we express all the
involved physical quantities in terms of functions of the scalar
field and their higher derivatives with respect to the scalar
field, by making simple assumptions, mainly the slow-roll
assumption for the scalar field and the slow-roll assumption
$\dot{H}\ll H^2$ which actually makes inflation possible to occur.
Indeed, if doing so, the gravitational equations of motion, the
slow-roll indices and the resulting observational indices have
quite simple and elegant final expressions, and the
phenomenological implications can be investigated in a much more
transparent and simple way, in comparison to our previous approach
\cite{Odintsov:2019clh}. Thus with the present paper, we would
like to present an elegant theory, with simple expressions in
closed form for the physical quantities involved, that may be
added in the already successful theories of modified gravity
\cite{Nojiri:2017ncd,Nojiri:2010wj,Nojiri:2006ri,Capozziello:2011et,Capozziello:2010zz,delaCruzDombriz:2012xy,Olmo:2011uz},
which are also compatible with the GW170817.

Our strategy to approach the GW170817-compatible
Einstein-Gauss-Bonnet inflationary theory is mainly based on the
imposed condition $c_T^2\simeq 1$, which results to the
differential equation $\ddot\xi-H\dot\xi=0$. We shall express the
latter in terms of the scalar field and functions of the scalar
field and their derivatives. By assuming that the slow-roll
conditions hold true for the scalar field and also for the Hubble
rate, we express the gravitational equations of motion in terms of
the scalar field, and also we calculate the slow-roll indices and
the observational indices as functions of the scalar field. One
important outcome of our theoretical framework is that the
Gauss-Bonnet scalar coupling function $\xi (\phi)$ and the scalar
potential $V(\phi)$, are strongly related to each other, a
condition that constrains the allowed functional forms of both
$\xi (\phi)$ and $V(\phi)$. With regard to the observational
indices, we are interested mainly in the spectral indices of the
scalar and tensor perturbations $n_{\mathcal{S}}$ and $n_T$
respectively, and the tensor-to-scalar ratio $r$. Thus we provide
a transparent theoretical framework with mathematically elegant
and simple expressions, that may be directly put to the test with
regard to its inflationary phenomenology implications. By choosing
several models of interest, we can express all the involved
quantities as functions of the $e$-foldings number, and the free
parameters for each model, and each model can be directly
confronted with the latest Planck (2018) constraints on inflation
\cite{Akrami:2018odb}. As we demonstrate, there exist several
models that can achieve viability with the observational data,
while at the same time they succeed to satisfy all the assumptions
made for deriving the equations of motion, such as the slow-roll
assumptions and so on. Finally, we examine the implications of one
further assumption well motivated by the slow-roll conditions,
namely $\xi'/\xi''\ll 1$. As we show, this constraint can also
lead to viable GW170817-compatible Einstein-Gauss-Bonnet
inflationary theories, which in fact are functionally more simple
in comparison to the previous case, where the constraint
$\xi'/\xi''\ll 1$ was not imposed. We also examine several models
of interest for this case, and we discuss several theoretical
implications of this theoretical framework.

\section{Einstein-Gauss-Bonnet Theories and GW170817 Compatibility Modifications}

We shall consider an Einstein-Gauss-Bonnet theory, which is
described by the following gravitational action,
\begin{equation}
\label{action} \centering
S=\int{d^4x\sqrt{-g}\left(\frac{R}{2\kappa^2}-\frac{\omega}{2}\partial_{\mu}\phi\partial^{\mu}\phi-V(\phi)-\frac{1}{2}\xi(\phi)\mathcal{G}\right)}\,
,
\end{equation}
where $R$ denotes the Ricci scalar, $\kappa=\frac{1}{M_p}$ with
$M_p$ being the reduced Planck mass, V($\phi$) is the scalar
potential, $\xi(\phi)$ is the Gauss-Bonnet coupling which is a
dimensionless function of the scalar field. Lastly, $\mathcal{G}$ is the
Gauss-Bonnet invariant in four dimensions, which is a scalar
quantity with dimensions $[m]^4$, with
$\mathcal{G}=R^2-R_{\alpha\beta}R^{\alpha\beta}+R_{\alpha\beta\gamma\delta}R^{\alpha\beta\gamma\delta}$
where $R_{\alpha\beta}$ and $R_{\alpha\beta\gamma\delta}$ being
the Ricci and Riemann tensor respectively.

It is worth mentioning that even though the gravitational action
involves $\omega$, which we assume to be just a constant, with
allowed values $\omega=\pm1$, our study will focus only on the
canonical case $\omega=1$, but we shall leave it as $\omega$ in
the equations that follow, in order to have the phantom scalar
case available. Nevertheless, as we mentioned, we shall focus on
the canonical scalar case. Furthermore, the cosmological
background will be assumed to be that of a flat spacetime with
Friedman-Robertson-Walker (FRW) metric, with the line element
being,
\begin{equation}
\label{metric} \centering
ds^2=-dt^2+a(t)^2\sum_{i=1}^{3}{(dx^{i})^2}\, ,
\end{equation}
where $a(t)$ denotes the scale factor. In addition, the scalar
field shall be assumed to be time-dependent only. Furthermore, the
Gauss-Bonnet scalar for the FRW metric is equal to
$\mathcal{G}=24H^2(\dot H+H^2)$.

By varying the gravitational action with respect to the metric
tensor and with respect to the scalar field, the gravitational
equations of motion are derived, which read,
\begin{equation}
\label{motion1} \centering
\frac{3H^2}{\kappa^2}=\frac{1}{2}\omega\dot\phi^2+V+12\dot\xi
H^3\, ,
\end{equation}
\begin{equation}
\label{motion2} \centering \frac{2\dot
H}{\kappa^2}=-\omega\dot\phi^2+4\ddot\xi H^2+8\dot\xi H\dot
H-4\dot\xi H^3\, ,
\end{equation}
\begin{equation}
\label{motion3} \centering
\omega(\ddot\phi+3H\dot\phi)+V'+12\xi'H^2(\dot H+H^2)=0\, .
\end{equation}
In order to study the dynamics of inflation, one needs an explicit
expression of Hubble's parameter and of the scalar field, by
solving the differential equations presented above. However, such
a system of differential equations is very difficult to solve
analytically and certain approximations must be made in order to
make it solvable. One usual and important assumption we shall made
is the slow-roll assumption,
\begin{equation}\label{slowrollhubble}
\dot{H}\ll H^2\, ,
\end{equation}
which is an essential assumption for the inflationary era to be
realized in the first place, and another assumption is that the
scalar field evolves in a slow-roll way, so the following usual
relations hold true,
\begin{equation}\label{scalarfieldslowroll}
\frac{\dot\phi^2}{2} \ll V,\,\,\,\ddot\phi\ll 3 H\dot\phi\, .
\end{equation}
Now let us get to the core of this article, the compatibility with
the observational data coming from the gravitational wave emission
of the event GW170817. As we already mentioned in the
introduction, the gravitational wave speed in natural units for
Einstein-Gauss-Bonnet theories has the form,
\begin{equation}
\label{GW} \centering c_T^2=1-\frac{Q_f}{2Q_t}\, ,
\end{equation}
where $Q_f=8(\ddot\xi-H\dot\xi)$, $Q_t=F+\frac{Q_b}{2}$,
$F=\frac{1}{\kappa^2}$ and $Q_b=-8\dot\xi H$. Hence,
compatibility may be achieved by equating the velocity of
gravitational waves with unity, or making it infinitesimally close
to unity. In other words, we demand $Q_f=0$ or $Q_f\simeq 0$. This
constraint leads to an ordinary differential equation
$\ddot\xi=H\dot\xi$. However, instead of solving this particular
differential equation, as was performed in a previous work of ours
\cite{Odintsov:2019clh}, we shall express it in terms of the
derivatives of the scalar field, so every function shall be
expressed in terms of the scalar field. Since
$\dot\xi=\xi'\dot\phi$ and $\frac{d}{dt}=\dot\phi\frac{d}{d\phi}$,
the differential equation can be written as,
\begin{equation}
\label{constraint1} \centering
\xi''\dot\phi^2+\xi'\ddot\phi=H\xi'\dot\phi\, .
\end{equation}
This equation is exactly equivalent to the differential equation
derived from the constrain $Q_f=0$. Assuming that,
\begin{equation}\label{firstslowroll}
 \xi'\ddot\phi \ll\xi''\dot\phi^2\, ,
\end{equation}
which is motivated from the slow-roll assumption of the scalar
field, Eq. (\ref{constraint1}) is greatly simplified and can be
solved with respect to the derivative of the scalar field,
\begin{equation}
\label{constraint} \centering
\dot{\phi}\simeq\frac{H\xi'}{\xi''}\, .
\end{equation}
As it is obvious by looking Eqs. (\ref{motion3}) and
(\ref{constraint}), the scalar field must obey both Eqs.
(\ref{motion3}) and (\ref{constraint}). Thus, we can rewrite the
third gravitational equation of motion Eq. (\ref{motion3}) with
respect to the Gauss-Bonnet scalar coupling function, as follows,
\begin{equation}
\label{motion4} \centering
\frac{\xi'}{\xi''}\simeq-\frac{1}{3\omega
H^2}\left(V'+12\xi'H^4\right)\, ,
\end{equation}
where we used the slow-roll assumption of Eq.
(\ref{slowrollhubble}). Furthermore we shall assume that the
additional following condition holds true,
\begin{equation}\label{scalarfieldslowrollextra}
12\dot\xi H^3=12\frac{\xi'^2H^4}{\xi''}\ll V\, ,
\end{equation}
so in view of Eqs. (\ref{slowrollhubble}),
(\ref{scalarfieldslowroll}), (\ref{constraint}) and
(\ref{scalarfieldslowrollextra}), the gravitational equations of
motion can be written in a very simplified form, as shown below,
\begin{equation}
\label{motion5} \centering H^2\simeq\frac{\kappa^2V}{3}\, ,
\end{equation}
\begin{equation}
\label{motion6} \centering \dot
H\simeq-\frac{1}{2}\kappa^2\omega\dot\phi^2\, ,
\end{equation}
\begin{equation}
\label{motion8} \centering \dot\phi\simeq\frac{H\xi'}{\xi''}\, .
\end{equation}
Also the combination of Eqs. (\ref{motion4}) and (\ref{motion5})
results in the following differential equation,
\begin{equation}
\label{maindiffeqn} \centering \frac{\xi'}{\xi''}=-\frac{1}{\omega
\kappa^2V}\Big{(}V' +\frac{4}{3}\xi'V^2 \kappa^4\Big{)}\, ,
\end{equation}
which must be obeyed by both the scalar coupling function $\xi
(\phi)$ and the scalar potential, and essentially it is very
important for the analysis that follows.

The equations (\ref{motion5}), (\ref{motion6}), (\ref{motion8}),
and (\ref{maindiffeqn}) show that in our approach, all the
quantities involved in the inflationary phenomenology of the
GW170817 compatible Einstein-Gauss-Bonnet model, can be expressed
as functions of the scalar field. This is very important, however,
the most appealing feature of our approach is the simplicity of
the slow-roll indices as functions of the scalar field. Let us
demonstrate this by directly calculating the slow-roll indices, in
view of Eqs.  (\ref{motion5}), (\ref{motion6}), (\ref{motion8}),
and (\ref{maindiffeqn}). The slow-roll indices for the theory at
hand are defined to be \cite{Hwang:2005hb},
\begin{align}\label{slowrollbasic}
\centering \epsilon_1&=-\frac{\dot H}{H^2},
&\epsilon_2&=\frac{\ddot\phi}{H\dot\phi}, & \epsilon_3&=\frac{\dot
F}{2HF}, & \epsilon_4&=\frac{\dot E}{2HE}, &
\epsilon_5&=\frac{\dot F+Q_a}{2HQ_t}, &\epsilon_6&=\frac{\dot
Q_t}{2HQ_t}\, ,
\end{align}
where $F=\frac{1}{\kappa^2}$ in the case at hand, and the function
$E$ is defined to be,
\begin{equation}\label{functionE}
E=\frac{F}{\dot\phi^2}\left(\omega\phi^2+3\left(\frac{(\dot
F+Q_a)^2}{2Q_t}\right)+Q_c\right)\, ,
\end{equation}
while the functions $Q_a$, $Q_t$, $Q_b$ and $Q_c$, and
additionally the function $Q_e$ are equal to \cite{Hwang:2005hb},
\begin{equation}\label{qis}
Q_a=-4\dot\xi H^2,\,\,\,Q_b=-8\dot\xi
H,\,\,\,Q_t=F+\frac{Q_b}{2},\,\,\,Q_c=0,\,\,\,Q_e=-16
\dot{\xi}\dot{H}\, ,
\end{equation}
and are characteristic contribution of the Gauss-Bonnet related
term to the dynamics of inflation. By using Eqs.
(\ref{motion5})-(\ref{motion8}), the functions $Q_i$ of Eq.
(\ref{qis}) can be expressed as functions of the scalar field, so
we quote here the resulting expressions, to be used in the
following,
\begin{equation}\label{Qa}
\centering Q_a\simeq-4\frac{\xi'^2}{\xi''}H^3\simeq
-\frac{\left(4  \kappa ^3\right) V(\phi )^{3/2} \xi '(\phi
)^2}{\left(3 \sqrt{3}\right) \xi ''(\phi )}\, ,
\end{equation}
\begin{equation}\label{Qb}
\centering Q_b\simeq-8\frac{\xi'^2}{\xi''}H^2\simeq
-\frac{\left(8  \kappa ^2\right) V(\phi ) \xi '(\phi )^2}{3 \xi
''(\phi )}\, ,
\end{equation}
\begin{equation}\label{Qe}
\centering Q_e\simeq8k^2\omega\frac{\xi'^4}{\xi''^3}H^3\simeq
\frac{V(\phi )^{3/2} \left(8  \kappa ^5 \omega \right) \xi
'(\phi )^4}{\left(3 \sqrt{3}\right) \xi ''(\phi )^3}\, .
\end{equation}
Moreover, we can also express the slow-roll indices of Eq.
(\ref{slowrollbasic}) as functions of the scalar field, and these
are,
\begin{equation}
\label{index1} \centering
\epsilon_1\simeq\frac{\kappa^2\omega}{2}\left(\frac{\xi'}{\xi''}\right)^2\,
,
\end{equation}
\begin{equation}
\label{index2} \centering
\epsilon_2\simeq-\epsilon_1+1-\frac{\xi'\xi'''}{\xi''^2}\, ,
\end{equation}
\begin{equation}
\label{index3} \centering \epsilon_3=0\, ,
\end{equation}
\begin{equation}
\label{index4} \centering
\epsilon_4\simeq\frac{\xi'}{2\xi''}\frac{E'}{E}\, ,
\end{equation}
\begin{equation}
\label{index5} \centering
\epsilon_5\simeq-\frac{2\kappa^4\xi'^2V}{3\xi''-4\kappa^4\xi'^2V}\,
,
\end{equation}
\begin{equation}
\label{index6} \centering
\epsilon_6\simeq-\frac{2\kappa^4\xi'^2V\left(1-\frac{1}{2}\kappa^2\omega\left(\frac{\xi'}{\xi''}\right)^2\right)}{3\xi''-4\kappa^4V\xi'^2}\,
,
\end{equation}
and the explicit form of the function $E(\phi)$ is,
\begin{equation}\label{functionE}
E(\phi)=\frac{\omega}{\kappa^2}+\frac{8^2\kappa^4\xi'^2V^2\xi''}{3\xi''\left(1-\frac{4\kappa^4\xi'^2V}{3\xi''}\right)}\,
.
\end{equation}
Now let us proceed to the observable quantities, which can be
expressed in terms of the slow-roll indices. We start off with the
spectral index of the scalar curvature perturbations and the
spectral index of the tensor perturbations, which in terms of the
slow-roll indices are \cite{Hwang:2005hb},
\begin{equation}
\label{spectralindex} \centering
n_{\mathcal{S}}=1-2\frac{2\epsilon_1+\epsilon_2+\epsilon_4}{1-\epsilon_1}\,
,
\end{equation}
\begin{equation}\label{tensorspectralindex}
n_T=-2\frac{\epsilon_1+\epsilon_6}{1-\epsilon_1}\, ,
\end{equation}
while the tensor-to-scalar ratio is defined to be
\cite{Hwang:2005hb},
\begin{equation}\label{tensortoscalar}
r=16\left|\left(\frac{\kappa^2Q_e}{4H}-\epsilon_1\right)\frac{2c_A^3}{2+\kappa^2Q_b}\right|\,
,
\end{equation}
with $c_A$ being the sound speed, which is equal to,
\begin{equation}
\label{sound} \centering
c_A^2=1+\frac{Q_aQ_e}{3Q_a^2+\omega\dot\phi^2(\frac{2}{\kappa^2}+Q_b)}\,
,
\end{equation}
for the Einstein-Gauss-Bonnet theory at hand. Finally, we can also
express the $e$-foldings number in terms of the scalar field as
well. By using definition, $N=\int_{t_i}^{t_f}{Hdt}$, where $t_i$
and $t_f$ signify the time instance at first horizon crossing  and
at the end of inflation respectively, and according to Eq.
(\ref{motion8}), the $e$-foldings number can be written as an
integral with respect to the scalar field, as follows,
\begin{equation}
\label{efolds} \centering
N=\int_{t_i}^{t_f}{Hdt}=\int_{\phi_i}^{\phi_f}\frac{H}{\dot{\phi}}d\phi=\int_{\phi_i}^{\phi_f}{\frac{\xi''}{\xi'}d\phi}\,
,
\end{equation}
where $\phi_i$ and $\phi_f$ are the values of the scalar field at
the first horizon crossing and at the end of the inflationary era
respectively. This is the final piece needed in order to extract
the phenomenological implications of the GW170817 compatible
Einstein-Gauss-Bonnet theory.

The strategy to explicitly check the phenomenological viability of
the GW170817 compatible Einstein-Gauss-Bonnet theory is the
following: Firstly we choose an appropriate functional form for
the scalar Gauss-Bonnet coupling $\xi (\phi)$, then by inserting
it in the differential equation (\ref{maindiffeqn}), the scalar
potential can be obtained. Accordingly, for these functions, the
slow-roll indices (\ref{index1})-(\ref{index6}) can be obtained as
functions of the scalar field. Then, we can evaluate the final
value of the scalar field at the end of the inflationary era by
equating $\epsilon_1\simeq 1$, and also by using the resulting
$\phi_f$, and after performing the integral (\ref{efolds}), we can
solve the resulting equation with respect to $\phi_i$, which
recall is the value of the scalar field at the first horizon
crossing, now evaluated as a function of the $e$-foldings number
and of the free parameters of each model. Finally by substituting
the value $\phi_i$ in the slow-roll indices
(\ref{index1})-(\ref{index6}), since these must be evaluated at
the first horizon crossing, we can obtain the slow-roll indices
(\ref{index1})-(\ref{index6}) and the observational indices
(\ref{spectralindex}), (\ref{tensorspectralindex}) and
(\ref{tensortoscalar}) as functions of the $e$-foldings number and
of the free parameters of each model. Finally, the resulting
expressions can be directly compared with the latest Planck data
(2018) \cite{Akrami:2018odb}, which constrain the spectral index
of the scalar perturbations $n_{\mathcal{S}}$ and the
tensor-to-scalar ratio $r$ as follows,
\begin{equation}\label{planck2018}
\centering n_{\mathcal{S}}=0.9649\pm0.0042,\,\,\, r<0.064
\end{equation}
With regards to the spectral index of the tensor perturbations,
there is no reason for the consistency relation of the canonical
scalar theory to hold true, so we just quote the value, and we do
not pursuit this issue further for the various models we shall
examine in the following sections.

In the next section we shall examine several models that can yield
a viable phenomenology in the context of the GW170817-compatible
Einstein-Gauss-Bonnet theory.

The choice for the Gauss-Bonnet coupling $\xi(\phi )$ which will
be done in the next sections, might seem bizarre in each case,
however there is a strategy we used in each case, and it is based
on the simplicity of the fractions $\xi'/\xi''$ and $\xi''/\xi'$.
Note that the fraction $\xi''/\xi'$ appears in the expression of
the $e$-foldings number integral (\ref{efolds}), so if an
appropriate choice for $\xi$ is made, the integral of the
$e$-foldings number (\ref{efolds}), can be performed easily.
Accordingly, the fraction $\xi'/\xi''$ appears in the differential
equation (\ref{maindiffeqn}), a suitable choice for $\xi'/\xi''$
may result to a simple form of the differential equation
(\ref{maindiffeqn}), and thus the scalar potential can easily be
obtained by solving it analytically. A not suitable choice of the
coupling function $\xi (\phi)$ would make the differential
equation (\ref{maindiffeqn}) unsolvable, at least analytically,
but the analyticity of the equations is our main target behind the
various choices of the coupling function. Another reason the
functional simplicity of the first slow-roll index $\epsilon_1$
(\ref{index1}), and in the Appendix we further discuss this issue
by using illustrative examples.

\section{Confronting the GW170817-compatible Einstein-Gauss-Bonnet Theory with Observations}

In this section we shall study explicit examples of GW170817
compatible Einstein-Gauss-Bonnet models that can yield a
phenomenologically viable inflationary era. Recall that the most
severe constraint is that the scalar coupling function $\xi(\phi)$
and the scalar potential $V(\phi)$ must satisfy the differential
equation (\ref{maindiffeqn}). The most easy way is to assume a
specific form for the function $\xi(\phi)$ and then solve the
differential equation (\ref{maindiffeqn}) and find the scalar
potential $V(\phi)$, and accordingly the resulting model can be
tested directly. However, most usual choices for the function $\xi
(\phi)$, like simple power-law models or combinations of
exponentials or even simple sinusoidal functions, to do not lead
to viable phenomenologies. We found some examples from which a
viable phenomenology can be obtained, but in principle
combinations of simple functions can also be tested.

\subsection{Model I: The Error Function Choice for $\xi(\phi)$}

A particularly interesting model with optimal viability properties
is obtained if we choose the coupling scalar function $\xi(\phi)$
to be equal to,
\begin{equation}
\label{modelA} \centering
\xi(\phi)=y_0\mathrm{Erf}(\gamma\kappa\phi)=\frac{2y_0}{\sqrt{\pi}}\int_{0}^{\gamma\kappa\phi}{e^{-x^2}dx}\,
,
\end{equation}
where $x$ is an auxiliary integration variable and $\gamma$, $y_0$
are dimensionless constants to be specified later on in this
subsection, and $\mathrm{Erf}(z)$ is the error function. At first
glance, this designation of the coupling function might seem odd.
Despite its appearance, this function has certain characteristics
which make it interesting. Firstly, the derivatives of such
function are connected via a simple but elegant equation,
\begin{equation}
\label{derA} \centering \xi''=-2\gamma^2\kappa^2\phi\xi'\, ,
\end{equation}
Subsequently, the slow-roll indices $\epsilon_1$, $\epsilon_2$,
the $e$-foldings number and the necessary scalar field values
$\phi_i$, $\phi_f$ have quite simple functional forms in their
final forms. Solving the differential equation for the scalar
potential in Eq. (\ref{maindiffeqn}), one finds that the resulting
solution for the scalar potential has the following form,
\begin{equation}
\label{potA} \centering V(\phi)=\frac{3 \sqrt{\pi } \phi
^{\frac{\omega }{2 \gamma ^2}} \left(\gamma ^2 \kappa ^2 \phi
^2\right)^{\frac{\omega }{4 \gamma ^2}+\frac{1}{2}}}{3 \sqrt{\pi }
\text{$c$} \left(\gamma ^2 \kappa ^2 \phi
^2\right)^{\frac{\omega }{4 \gamma ^2}+\frac{1}{2}}-4 \gamma y_0
 \kappa ^5 \phi ^{\frac{\omega }{2 \gamma ^2}+1}
\Gamma \left(\frac{1}{4} \left(\frac{\omega }{\gamma
^2}+2\right),\gamma ^2 \kappa ^2 \phi ^2\right)}\, ,
\end{equation}
where $c$ is an arbitrary  integration constant with mass
dimensions [m]$^{\frac{\omega}{2\gamma^2}-4}$. Instead of naively
equating it to zero, we shall keep it and examine whether in can
be of any use. Similar to the scalar potential, the slow-roll
indices can be evaluated using the coupling scalar function in Eq.
(\ref{modelA}), as shown below,
\begin{equation}
\label{index1A} \centering
\epsilon_1\simeq\frac{\omega}{8(\gamma^2\kappa\phi)^2} \, ,
\end{equation}
\begin{equation}
\label{index2A} \centering
\epsilon_2\simeq\frac{\omega-4\gamma^2}{8(\gamma^2\kappa\phi)^2}\,
,
\end{equation}
\begin{equation}
\label{index3A} \centering \epsilon_3=0\, ,
\end{equation}
\begin{equation}
\label{index5A} \centering \epsilon_5\simeq\frac{2
\kappa ^3 \text{$y_0$} \phi ^{\frac{\omega }{2 \gamma ^2}}}{4
\kappa ^3 \phi ^{\frac{\omega }{2 \gamma ^2}} \left(
\text{$y_0$}-\gamma ^2 \text{$y_0$} \kappa ^2 \phi ^2 e^{\gamma
^2 \kappa ^2 \phi ^2} E_{\frac{1}{2}-\frac{\omega }{4 \gamma
^2}}\left(\gamma ^2 \kappa ^2 \phi ^2\right)\right)+3 \sqrt{\pi }
\gamma  \text{$c$} \phi e^{\gamma ^2 \kappa ^2 \phi ^2}}\, ,
\end{equation}
\begin{equation}
\label{index6A} \centering \epsilon_6\simeq\frac{ \kappa
\left(\omega -8 \gamma ^4 (\kappa \phi) ^2\right)}{4 \gamma ^4
\phi ^2 \left(4 \gamma ^2 \kappa ^5 \phi ^2 e^{\gamma ^2
\kappa ^2 \phi ^2} E_{\frac{1}{2}-\frac{\omega }{4 \gamma
^2}}\left(\gamma ^2 \kappa ^2 \phi ^2\right)-4 \kappa ^3-3
\sqrt{\pi } \gamma  c e^{\gamma ^2 \kappa ^2 \phi ^2} \phi
^{1-\frac{\omega }{2 \gamma ^2}}\right)}\, ,
\end{equation}
where $E_x$ denotes the exponential integral with lower limit x,
and we omitted the slow-roll index  $\epsilon_4$ because it has a
quite lengthy final expression. In addition, the first two indices
are described by simple and elegant equations compared to the
rest, which is expected since they are connected to the
derivatives of the coupling scalar function $\xi(\phi)$.
\begin{figure}
\includegraphics[width=20pc]{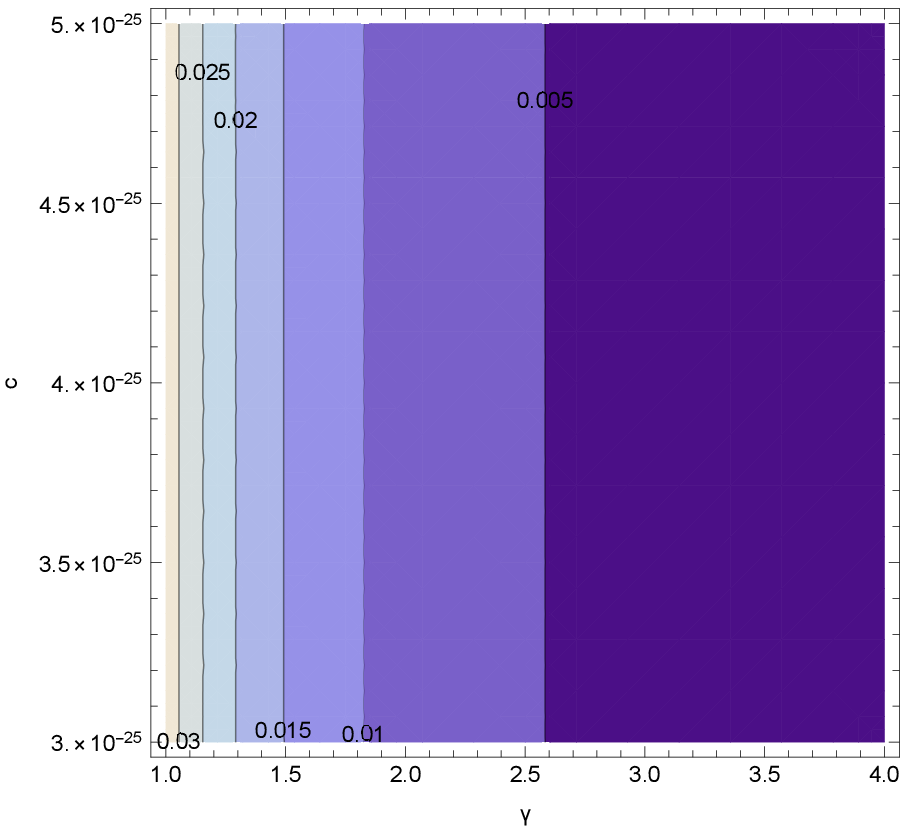}
\includegraphics[width=20pc]{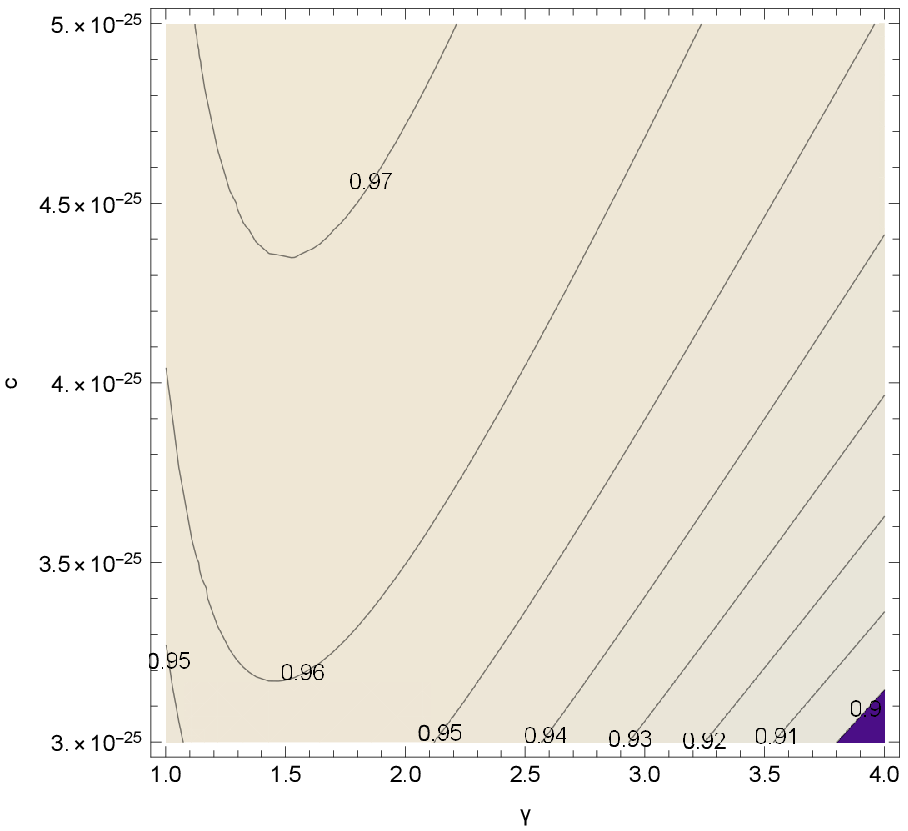}
\caption{ Contour plots of the tensor-to-scalar ratio $r$ (left
plot) and the spectral index $n_{\mathcal{S}}$ (right plot)
depending on parameters $c$ and $\gamma$. Their values range
from [$3\times 10^{-25}$,$5\times 10^{-25}$] and [1,4] for $c$
and $\gamma$ respectively.} \label{plot1}
\end{figure}
This is exactly why the error function was chosen in the first
place. In consequence, we can evaluate the final value of the
scalar field by letting slow-roll index $\epsilon_1$ in Eq.
(\ref{index1A}) become equal to unity. Doing so, we end up with
two values for the scalar field,
\begin{equation}
\label{scalarfA} \centering
\phi_f=\pm\frac{\sqrt{\omega}}{2\sqrt{2}\kappa\gamma^2}\, ,
\end{equation}
Hence, recalling the $e$-foldings number formula in Eq.
(\ref{efolds}) and using the previous result, the initial value of
the scalar field at the first horizon crossing reads,
\begin{equation}
\label{scalariA}
\phi_i=\pm\frac{\sqrt{8N\gamma^2+\omega}}{2\sqrt{2}\kappa\gamma^2}\,
,
\end{equation}
Each value of the scalar field is given by two signs, either a
plus or a minus, but we keep the physically consistent value,
which is the positive of course. For simplicity, we shall use the
reduced Planck physical units system, for which $\kappa^2=1$.
Assuming that the free parameters have the following values
($\omega$, $y_0$, $\gamma$, $c$)=(1, 1, 2, 4.09413$\times
10^{-25}$) in reduced Planck units, meaning $\kappa=1$, the
spectral indices and the tensor-to-scalar ratio become equal to
$n_{\mathcal{S}}=0.966$, $n_T=-0.00342555$ and $r=0.00832892$,
which are compatible with the latest Planck data
\cite{Akrami:2018odb} of Eq. (\ref{planck2018}), at least when the
tensor-to-scalar ratio and the spectral index of scalar curvature
are considered. The maximum bound for the tensor spectral index
coming from Planck 2018 \cite{Akrami:2018odb} is $n_T\simeq 2$ so
the present model is also compatible with this constraint, however
the tensor tilt coming from the Planck data is strongly related to
the minimally coupled canonical scalar field consistency relation
assumption, so it is conceivable that the gravity of the tensor
tilt $n_T$ result cannot be taken into account as seriously as the
spectral index and the tensor-to-scalar ratio. Furthermore, the
value of the sound speed for the above values of the parameters is
$c_A^2\simeq 0.999994$, so the theory is free from instabilities.
Additionally, the values of the scalar field are $\phi_i=3.87399$
and $\phi_f=0.0883883$ and due to continuity, in can easily be
inferred that the value of the scalar field decreases with time.
\begin{figure}
\includegraphics[width=20pc]{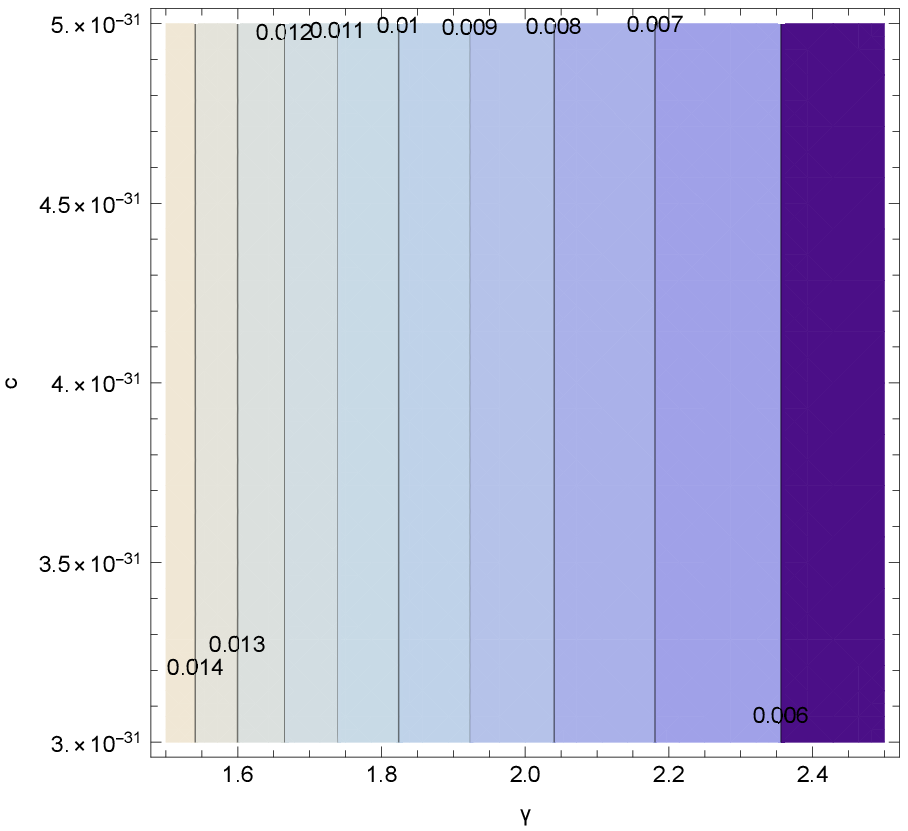}
\includegraphics[width=20pc]{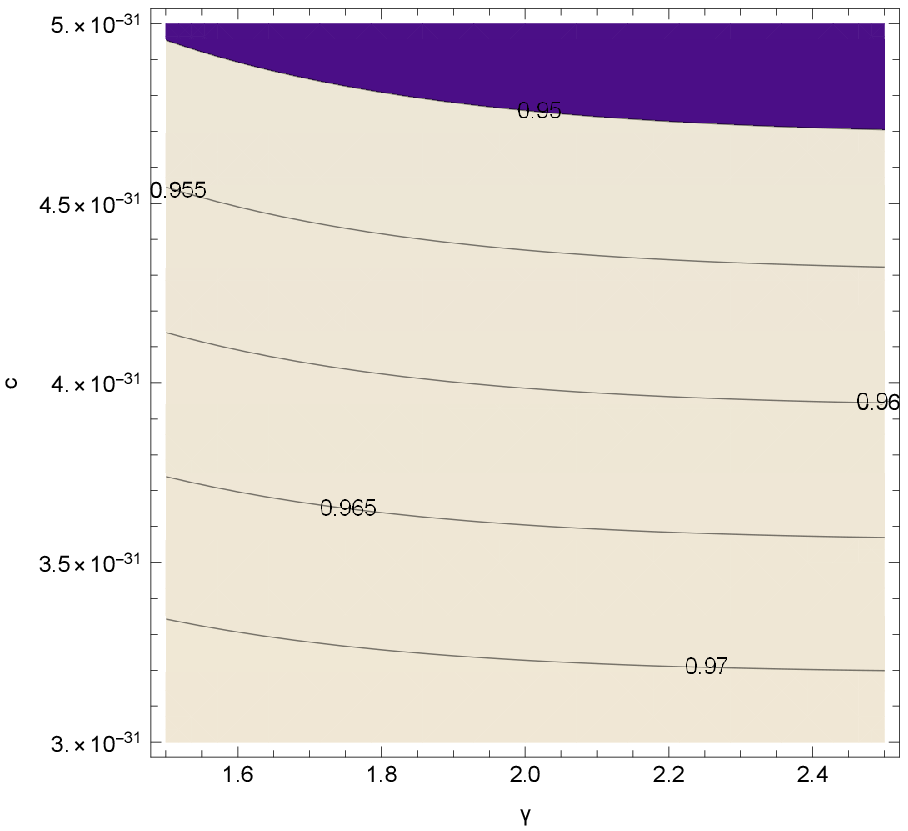}
\caption{Contour plots of the tensor-to-scalar ratio r (left) and
the spectral index $n_{\mathcal{S}}$ (right) depending on
parameters $c$ and $\gamma$. Their values range from [$3\times
10^{-31}$,$5\times 10^{-31}$] and [1.5,2.5] for $c$ and $\gamma$
respectively.} \label{plot2}
\end{figure}
It is worth mentioning that this particular set of parameters is
not the only one capable of producing viable inflation. It turns
out that there exist four different values for the integration
constant $c$ which yield the same values for the observed
quantities by keeping the rest parameters fixed. Apart from the
one used previously, inserting either one from the values in
reduced Planck units $c=-4.07962\times 10^{-25}$,
$c=3.52933\times 10^{-31}$ or $c=4.72821\times 10^{-28}$
produces the exact same result, implying that there exist more
possible values, therefore multiple viable parameter values
regions which could produce viable inflation. The following plots
depict such regions of viability for the parameters $c$ and
$\gamma$. It is obvious from Fig. \ref{plot1} and Fig. \ref{plot2}
that $\gamma$ affects mainly the spectral index $n_{\mathcal{S}}$
while $c$ both the spectral index and the tensor-to-scalar
ratio. For the sake of consistency, we mention that the choice for
such small integration constant in Planck-Units is in a way mandatory
in order to achieve compatibility with the Planck 2018 data and not a
aimed choice of ours. In fact, since the rest free parameters obtained
such values, the integration constant essentially was forced to obtain such
a small value.

Lastly, we must check whether our approximations we made in the
previous section are valid, for the values of the free parameters
for which the viability of the model when compared to the Planck
data is ensured. By choosing ($\omega$, $y_0$, $\gamma$,
$c$)=(1, 1, 2, 4.09413$\times 10^{-25}$) in reduced Planck
units, we have that $\frac{\dot H}{H^2}\sim\mathcal{O}(10^{-3})$,
so the slow-roll condition (\ref{slowrollhubble}) holds true.
Similarly, the kinetic term at the same epoch is of order
$\frac{1}{2}\omega\dot\phi^2\sim\mathcal{O}(10^{20})$ while the
scalar potential is V($\phi$)$\sim\mathcal{O}(10^{24})$,
therefore, the slow-roll approximation for the scalar field
(\ref{firstslowroll}) holds also true. In addition, let us check
the condition (\ref{scalarfieldslowrollextra}), namely,
$12\frac{\xi'^2H^4}{\xi''}\ll V$, so for ($\omega$,
$y_0$, $\gamma$, $c$)=(1, 1, 2, 4.09413$\times 10^{-25}$),
the fraction of the two terms, namely
$\frac{12\frac{\xi'^2H^4}{\xi''}}{V}$, is approximately
$\frac{12\frac{\xi'^2H^4}{\xi''}}{V}\sim \mathcal{O}(2\times
10^{-3})$, so the approximation is valid in this case too. In
conclusion, the error function choice for the scalar Gauss-Bonnet
coupling $\xi(\phi)$, yields a phenomenologically viable
inflationary era for the GW170817-compatible Einstein-Gauss-Bonnet
model.
\begin{figure}
\includegraphics[width=20pc]{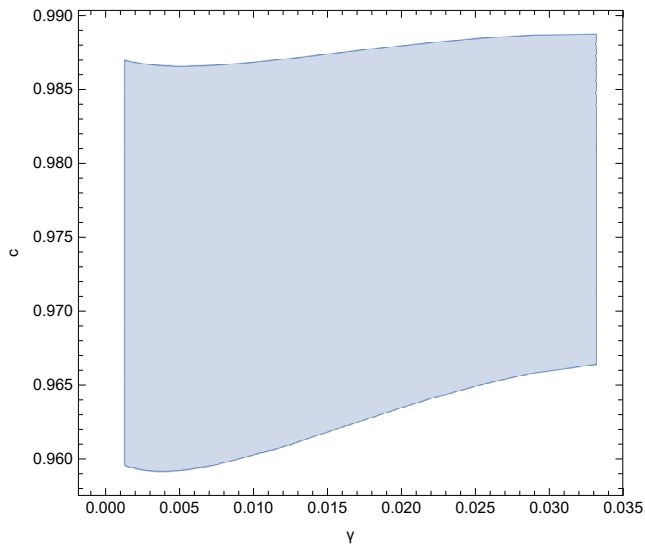}
\caption{Parametric plot of the tensor-to-scalar ratio r (x axis)
and the spectral index $n_{\mathcal{S}}$ (y axis) depending on
parameters $c$ and $\gamma$. Their values range from [$2\times
10^{-31}$,$4\times 10^{-31}$] and [1,5] for $c$ and $\gamma$
respectively.} \label{parplot1}
\end{figure}
Finally, it is easy to check that all the slow-roll indices
satisfy the relation $\epsilon_i \ll 1$, $i=1,2,4,5,6$. Indeed,
for ($\omega$, $y_0$, $\gamma$, $c$)=(1, 1, 2,
4.09413$\times 10^{-25}$) we have $\epsilon_1\simeq 0.000520562$,
$\epsilon_2\simeq 0.00780843$, $\epsilon_4\simeq 0.00814159$,
$\epsilon_5\simeq 0.00119194$ and $\epsilon_6\simeq -2.96645\times
10^{-25}$ at the first horizon crossing. Thus, this verifies that
the slow-roll condition indeed holds true.

Let us note here that the parameter $c$ has an extremely small
value compared to the rest free parameters of the model, even in
Planck units. This choice however was made in order to extract a
viable phenomenology for the specific model at hand namely model
(\ref{modelA}). In an essence, our analysis showed that the
parameter $c$ is forced to take such small and fine-tuned values,
in order for the rest of the parameters to have less fine-tuned
values, and simultaneously in order to obtain a viable
phenomenology. Perhaps, a complete different designation of the
free parameters value could lead to a viable phenomenology,
without such extreme fine-tuning on the parameter $c$. This case
is a possibility however we refrained from further analysis the
parameter space, because the model itself is just a choice made
for demonstrative reasons. It is obvious that a more refined model
would require less fine-tuning to the free parameters, and as we
show in the next sections, this is indeed the case.

\subsection{Phenomenology of a More Involved Model}

\begin{figure}
\includegraphics[width=20pc]{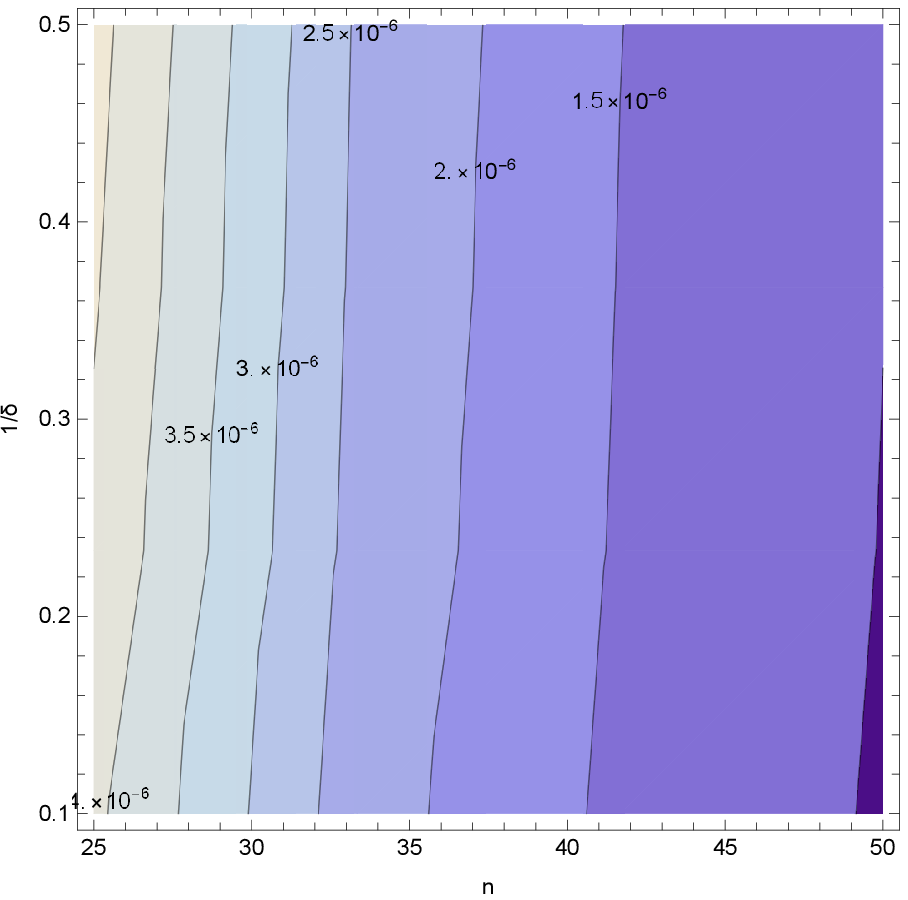}
\includegraphics[width=20pc]{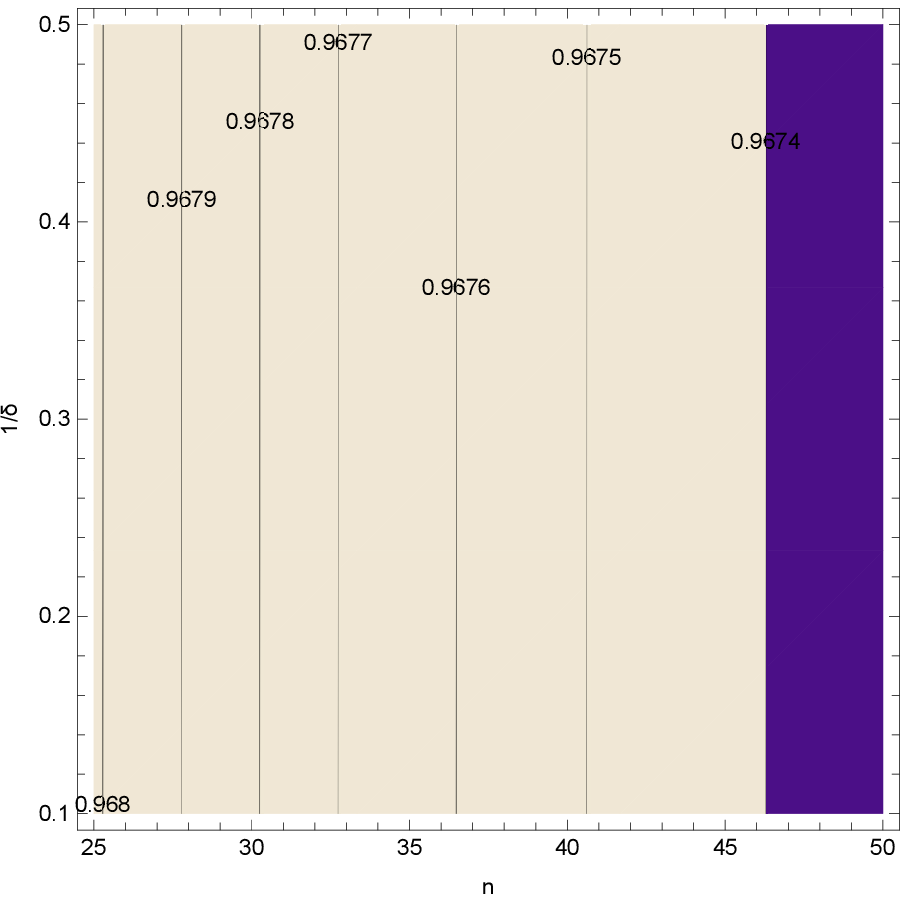}
\caption{Contour plots of the tensor-to-scalar ratio $r$ (left)
and the spectral index $n_{\mathcal{S}}$ (right) depending on the
dimensionless parameters $1/\delta$ and $n$. Their values range
from [$0.1$,$0.5$] and [25,50] for $1/\delta$ and $n$
respectively.} \label{plot3}
\end{figure}

\begin{figure}
\includegraphics[width=20pc]{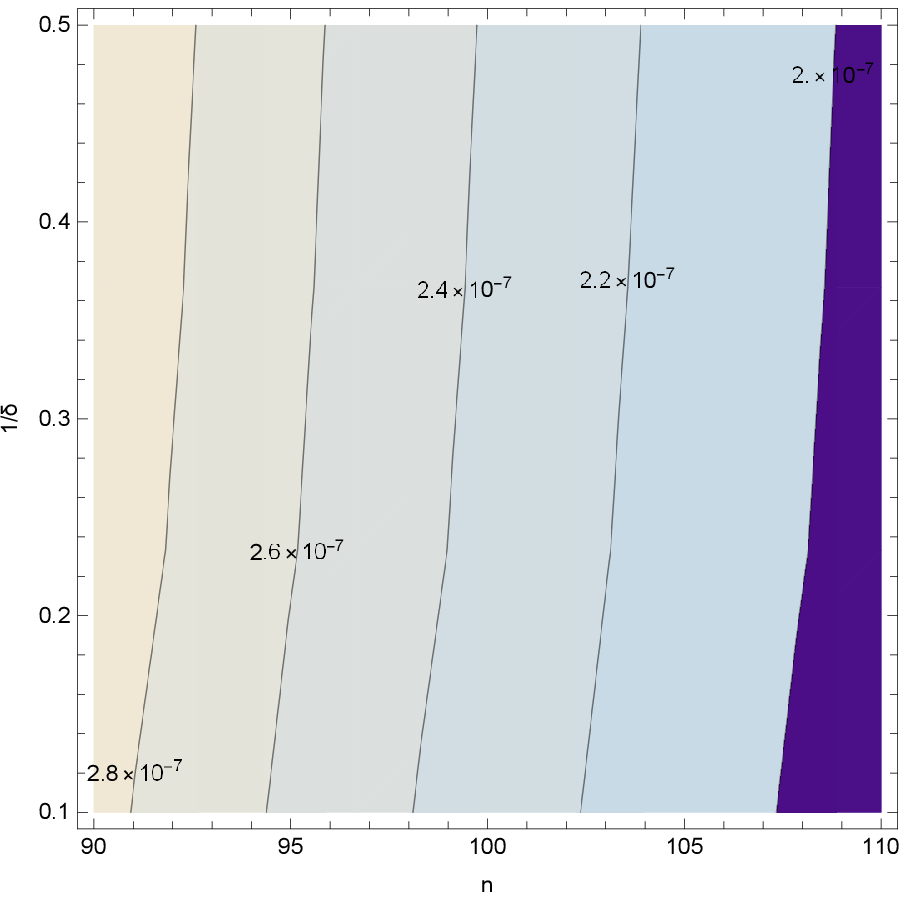}
\includegraphics[width=20pc]{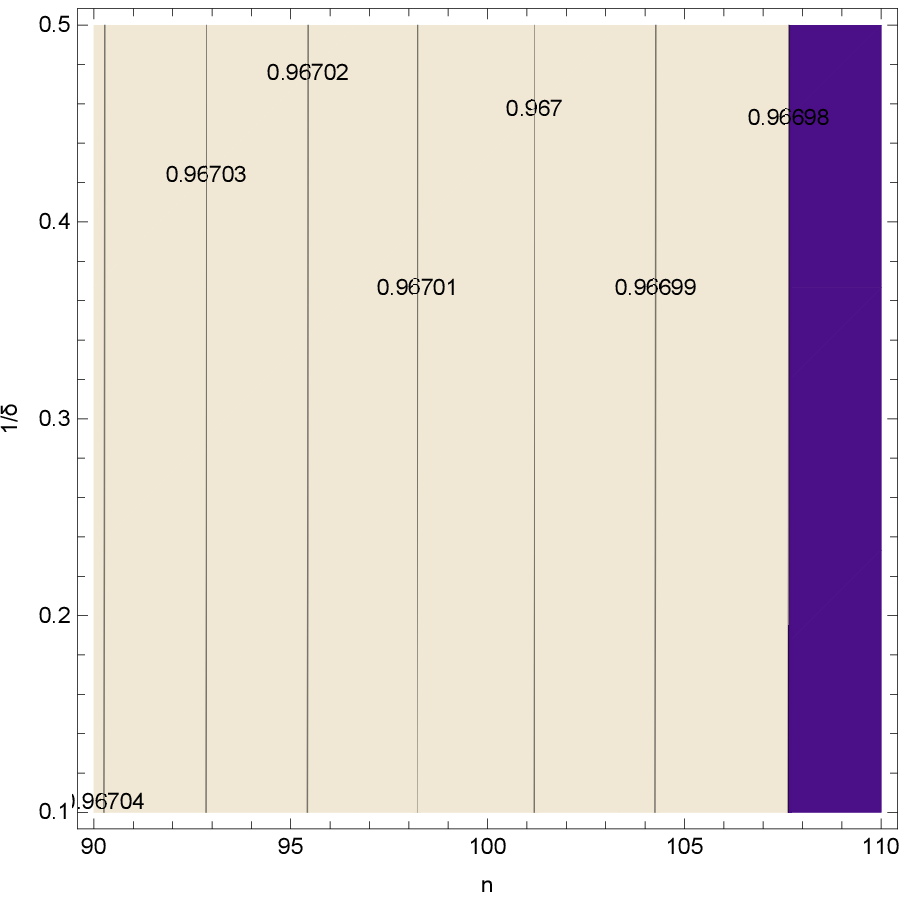}
\caption{Contour plots of the tensor-to-scalar ratio r (left) and
the spectral index $n_{\mathcal{S}}$ (right) depending on the
dimensionless parameters $1/\delta$ and $n$. Their values range
from [$0.1$,$0.5$] and [90,110] for $1/\delta$ and $n$
respectively.} \label{plot4}
\end{figure}

Suppose now that the Gauss-Bonnet coupling scalar function has the
following form,
\begin{equation}
\label{modelB} \centering
\xi(\phi)=\int^{\phi}{y_1e^{-\delta(\kappa\tau)^n}d\tau}\, ,
\end{equation}
where $y_1$, $\delta$ and $n$ are dimensionless constants to be
specified later. As it was the case with the previous model, the
second derivative of the coupling function is connected with the
first via a generalized equation compared to the first model,
\begin{equation}
\label{derB} \centering \xi''=-n\delta\kappa^n\phi^{n-1}\xi'\, .
\end{equation}
Thus, both the slow-roll indices $\epsilon_1$ and $\epsilon_2$,
the $e$-foldings number $N$ and the derivative of the scalar field
$\dot\phi$ are given again by simple expressions which are
proportional to the exponent $n$ of the model function
(\ref{modelB}). Consequently, specifying the exponent should in
principle produce expressions for the observable quantities
depending strongly on the exponent. On the other hand, the scalar
potential derived from Eq.(\ref{maindiffeqn}) has a complicated
form, as is shown below,
\begin{equation}
\label{potB} \centering V(\phi)=\frac{e^{\frac{\omega  (\kappa
\phi) ^{3-n}}{(3-n)\delta n}}}{\int_1^{\phi }
\frac{4}{3}\kappa^4 y_1 e^{\frac{k^{3-n} \omega
\tau^{3-n}}{(3-n)\delta n}-\delta\kappa^n\tau^n} \, d\tau}\, ,
\end{equation}
and $\tau$ an auxiliary integration variable. Moreover, the
arbitrary constant derived from the integration is assumed to be
equal to zero. However, the scalar potential enters only in the
equations through the Hubble rate, so it will affect only the rest
of the slow-roll indices and the observable quantities as well.
Since the scalar potential is also depending on the exponent of
the Gauss-Bonnet scalar function, the dominant factor which in the
end shall determine the viability of the model is this exponent.
Let us now proceed with the evaluation of the slow-roll indices.
Recalling their previous definitions in Eq. (\ref{index1}) through
Eq. (\ref{index3}) and the coupling function $\xi(\phi)$ in Eq.
(\ref{modelB}) as well, we end up with the following formulas.
\begin{equation}
\label{index1B} \centering
\epsilon_1\simeq\frac{\omega}{2}\left(\frac{1}{n\delta(\kappa\phi)^{n-1}}\right)^2\,
,
\end{equation}
\begin{equation}
\label{index2B} \centering
\epsilon_2\simeq\frac{(n-1)}{n\delta(\kappa\phi)^n}-\epsilon_1\, ,
\end{equation}
\begin{equation}
\label{index3B} \centering \epsilon_3=0\, ,
\end{equation}
Apparently, the first three slow-roll indices have very simple
forms, while the rest were omitted due to their lengthy final
forms. Note however that the indices $\epsilon_4$ and $\epsilon_6$
participate in the evaluation of the spectral indices and the
tensor-to-scalar ratio directly. The values of both the spectral
indices and the tensor-to-scalar ratio, as mentioned before, can
be calculated by utilizing the slow-roll indices introduced
previously. Similarly, from Eq. (\ref{index1B}), the value of the
scalar field at the end of inflation can be derived by equating
the index $\epsilon_1$  with unity. Thus, the final value of the
scalar field in this case is,
\begin{equation}
\label{scalarfB} \centering
\phi_f=\frac{1}{\kappa}\left(\frac{1}{\delta
n}\sqrt{\frac{\omega}{2}}\right)^{\frac{1}{n-1}}\, ,
\end{equation}
Similarly, the initial value of the scalar field at the first
horizon crossing can be inferred from the final value and the
$e$-foldings number in Eq. (\ref{efolds}) by simply solving the
integral. Thus, the initial value is,
\begin{equation}
\label{scalariB} \centering \phi_i= \frac{1}{\kappa}\left(
\left(\frac{1}{n\delta}\sqrt{\frac{\omega}{2}}\right)^{\frac{n}{n-1}}+\frac{N}{\delta}\right)^{\frac{1}{n}}\,
,
\end{equation}
where $N\simeq60$. An observant reader might notice that the two
previous results are presented in an incomplete manner, since the
expression of the final value of the scalar field should produce
at least $2n-2$ solutions while the initial value, only $n$.
Mathematically speaking, that would be correct, however, in order
to avoid the emergence of complex numbers, it was deemed necessary
to choose the positive value in each case. These values in fact
will lead to a viable model while the rest are physically
inconsistent. Hence, the positive value at the initial stage of
inflation shall be used as input in the spectral index and
tensor-to-scalar ratio in order to calculate the observable
quantities and ascertain the validity of the model by comparing
them with the values obtained by the Planck 2018 collaboration
\cite{Akrami:2018odb}. Let us assume that in Planck Units, in
reduced Planck units with $\kappa^2=1$, the free parameters of the
theory have the following fixed values, ($\omega$, $y_1$,
$\delta$, n)=(1, 0.0001, 3.33, 100). According to the previous
results for the scalar field in Eq. (\ref{scalariB}) and Eq.
(\ref{scalarfB}), the initial and final value of the scalar field
becomes equal to $\phi_i=1.02933$ and $\phi_f=0.939715$
respectively in Planck Units. At first site, it is clear that the
scalar field again decreases with time. Consequently, the
observable quantities take the values $n_{\mathcal{S}}=0.967004$,
and $r=2.35395\times 10^{-7}$, which are both compatible with
latest Planck observations (\ref{planck2018}) and the unobserved
for now spectral index of the tensor perturbations is equal to
$n_T=-2.94244\times 10^{-8}$, which is incompatible with the upper
bound of the Planck data $n_T\sim 2$ \cite{Akrami:2018odb}, but
still the Planck result depends strongly on the consistency
relation for a minimally coupled canonical scalar theory.
Generally speaking, the previous results were obtained for
specific values of the free parameters and especially, for a fixed
value of the exponent $n$ in Eq. (\ref{modelB}). However, this is
not the only set of parameters capable of producing compatible
results with the observations. It seems that each parameter is
insignificant compared to the exponent $n$, with the latter having
a dominant effect on the phenomenology produced. However, there is
a wide range of values of the parameter $n$, which may range from
[15,120] and even further, and as $n$ increases, both the spectral
index of primordial curvature perturbations $n_{\mathcal{S}}$ and
the tensor-to-scalar ratio $r$ decrease, but at different rates.
Specifically, the rate of decrease for the spectral index is
lesser than the rate of decrease for the tensor-to-scalar ratio.
Since there exists no lower boundary for the latter, there exists
a wide range of possible values for the exponent $n$ as is shown
in Fig. \ref{plot3} and Fig. \ref{plot4}. In the plots we present
two cases for which the viability of the model is achieved as
mentioned before. Parameters $1/\delta$, for convenience, and $n$
were chosen to study the response of the model in such changes. In
contrast to the previous model, it seems that now, the
tensor-to-scalar ratio depends on $\delta$ while the spectral
index remains unaffected. In addition, the exponent as expected
affects strongly both quantities.
Lastly, we discuss the validity of the approximations made
throughout our calculations. Firstly, the slow-roll condition for
the scalar field (\ref{firstslowroll}), so by choosing ($\omega$,
 $y_1$, $\delta$, n)=(1, 0.0001, 3.33, 100) in our case
we have, $\frac{1}{2}\omega\dot\phi^2\sim \mathcal{O}(10^{-5})$
while $V(\phi)\sim\mathcal{O}(10^4)$ in reduced Planck units, so
apparently it holds true. Also, $\dot H\sim\mathcal{O}(10^{-5})$
and $H^2\sim\mathcal{O}(10^3)$, and therefore the condition
(\ref{slowrollhubble}) also holds true. Similarly, in the first
gravitational equation of motion, the term $\sim \dot\xi H^3$ was
omitted as it was deemed small compared to the value of the scalar
potential. This assumption is proven to be true since at the
horizon crossing, the order of magnitude of this term is much
smaller than the corresponding one for the scalar potential, since
$\dot\xi H^3\sim\mathcal{O}(10^{-31})$, while
$V(\phi)\sim\mathcal{O}(10^4)$ in reduced Planck units at the
horizon crossing. Finally, we note that the initial ratio of the
first two derivatives of the Gauss-Bonnet coupling scalar function
is of order $\xi'/\xi''\sim\mathcal{O}(10^{-8})$, yet again it is
something expected since this ratio is connected with the ratio
$\dot H/H^2$. In the next section we shall further discuss this
issue.
As a last comment, it is worth mentioning certain similarities,
and differences as well, between the two models. Setting $n=2$,
$\delta=4$ and $y_1=2$, in principle, the models should coincide
since the ratio $\xi'/\xi''$ is exactly the same. Following this
research line, the initial and the final values of the scalar
field are also the same, something that is expected since it
attributed to the previous ratio equivalence. However, since in
the second model,the integration constant is zero, in contrast to the first, the scalar
potential will be different. Therefore, the indices $\epsilon_4$
through $\epsilon_6$, the sound wave speed $c_A$ and the spectral
index $n_{\mathcal{S}}$ as well, are different in each model since
these parameters depend on the scalar potential. For the
tensor-to-scalar ratio which also depends from the sound wave
speed, the result is the same up some accuracy, implying that the
dominant factor, which in fact experiences greater change, is the
index $\epsilon_4$. Despite that, there is no limitation that
forbids the two models in agreeing with each other but if that is
the case, a different set of parameters is needed. Furthermore,
the previous analysis has made it abundantly clear that in order
to yield viable results, one can work in two separate ways. Either
freeze the exponent $n$ and find an appropriate initial condition
for the scalar potential, meaning designating properly the
arbitrary integration constant, or neglect this particular
constant by equating it with zero, and vary the exponent $n$.
Having both the exponent and the integration constant taking
values simultaneously is surely a possibility, that in principle
could yield viable results, but this issue is a by far more
complicated task.

\subsection{Phenomenology Under the Assumption  $\xi'/\xi''\ll 1$}

Let us consider again the condition $\frac{\dot{\phi}^2}{2}\ll V$,
which can be rewritten as,
\begin{equation}\label{constraintnewsec}
\frac{\frac{\dot{\phi}^2}{2}}{V}\ll 1\, ,
\end{equation}
and by using Eq. (\ref{motion8}), we can write Eq.
(\ref{constraintnewsec}) as follows,
\begin{equation}\label{conditionmain1}
\frac{\kappa^2}{6}\frac{\xi'^2}{\xi''^2}\ll 1\, .
\end{equation}
It is rather tempting to investigate the phenomenology of the
GW170817-compatible Einstein-Gauss-Bonnet model in the case that
the following additional condition holds true,
\begin{equation}\label{maincondition}
\kappa \frac{\xi'}{\xi''}\ll 1\, ,
\end{equation}
which is motivated from the condition
$\frac{\xi'^2\kappa^2}{6\xi''^2}\ll 1$ of Eq.
(\ref{conditionmain1}). In this case, in view of the constraint
(\ref{maincondition}) the term $\frac{\xi'}{\xi''}$ can be
disregarded in Eq. (\ref{maindiffeqn}), so the latter becomes,
\begin{equation}
\label{maindiffeqnslowroll} \centering \frac{V'}{V}
+\frac{4}{3}\xi'V \kappa^4\simeq 0\, .
\end{equation}
This means that the two terms $\sim \frac{V'}{V}$ and $\sim \xi'V$
are of the same order in Planck units. Also, it is obvious that in
the case at hand, the differential equation
(\ref{maindiffeqnslowroll}) that yields the scalar potential, for
a given function $\xi(\phi )$, or vice versa, is more easy to
solve analytically. Another motivation for choosing the condition
(\ref{maincondition}) is the fact that the first slow-roll index
$\epsilon_1$ in Eq. (\ref{index1}) is proportional to the ratio
$\left(\xi'/\xi''\right)^2$ and the value of such index at the
first horizon crossing is expected to be much lesser than unity,
if the slow-roll conditions apply in the theory. Thus, it stands
to reason why this ratio can be neglected.

In this section, we shall investigate the phenomenological
implications of the condition (\ref{maincondition}) on the
GW170817-compatible Einstein-Gauss-Bonnet theory. What will
actually change in the whole procedure we developed in the
previous section, is the relation that gives the scalar potential
$V(\phi)$ given the scalar coupling function $\xi (\phi)$ and vice
versa. The relations that yield the slow-roll indices as functions
of the scalar field and the corresponding observational indices
remain the same, so effectively we have a simpler theoretical
framework. Let us demonstrate how the phenomenology of the
GW170817-compatible Einstein-Gauss-Bonnet theory is modified in
view of the assumption (\ref{maindiffeqnslowroll}). In the
following, we shall examine three simple models and explicitly
confront the models whether they lead to viable results or
inconsistencies.

Suppose first that the coupling function is given by a simple
power-law scalar field dependence,
\begin{equation}
\label{powerlaw} \centering \xi(\phi)=\lambda(\kappa\phi)^m\, ,
\end{equation}
were $\lambda$ is a dimensionless constant. This particular model
also belongs to the same category as the previous ones, since
there exists a simple connection between the derivatives of
$\xi(\phi)$. Specifically,
\begin{equation}
\centering \xi''=\frac{(m-1)}{\phi}\xi'\, ,
\end{equation}
so accordingly, the slow-roll indices $\epsilon_1, \epsilon_2$,
the $e$-foldings number relation and the initial-final value of
the scalar field, are given by simple expressions. Following the
same procedure as in the previous section, the scalar potential
can be extracted from Eq. (\ref{maindiffeqnslowroll}), which
reads,
\begin{equation}
\label{V1} \centering
V(\phi)=\frac{1}{\frac{4}{3}\kappa^4\xi(\phi)-c_1}\, ,
\end{equation}
where $c_1$ is an arbitrary integration constant with mass
dimensions [m]$^{-4}$. This is a much simpler expression for the
scalar potential compared to the models of the previous sections.
Now we shall examine the viability of the power-law model where
the arbitrary integration constant in non-zero and accordingly we
shall examine the case when this particular constant is in fact
equal to zero. The latter is a very interesting case since as it
can be inferred from Eq. (\ref{V1}), the product of the scalar
potential and the Gauss-Bonnet coupling scalar function is a well
defined constant, and as a matter of fact, one with very
restricted form, as it can be inferred by Eq.
(\ref{maindiffeqnslowroll}). Let us proceed with the first case
where the the integration constant is nonzero. Then, similar to
the previous two models, the slow-roll indices derived from Eqs.
(\ref{index1})-(\ref{index6}) have the following form,
\begin{equation}
\label{indexC1} \centering \epsilon_1\simeq\frac{ \omega (\kappa
\phi) ^2}{2 (m-1)^2}\, ,
\end{equation}
\begin{equation}
\centering \epsilon_2\simeq-\frac{ \omega  (\kappa \phi) ^2-2
m+2}{2 (m-1)^2}\, ,
\end{equation}
\begin{equation}
\centering \epsilon_3=0\, ,
\end{equation}
\begin{equation}
\centering \epsilon_5\simeq\frac{2 \kappa ^4 \lambda  m
(\kappa  \phi )^m}{3 c_1 (m-1)+4 \kappa ^4 \lambda (\kappa
\phi) ^m}\, ,
\end{equation}
\begin{equation}
\centering \epsilon_6\simeq\frac{ \kappa ^4 \lambda  m (\kappa
\phi )^m \left(-\omega (\kappa \phi) ^2+2( m-1)^2\right)}{(m-1)^2
\left(3 c_1 (m-1)+4  \kappa ^4 \lambda  (\kappa  \phi
)^m\right)}\, .
\end{equation}
For this model, the slow-roll indices have a particularly simple
form, apart from $\epsilon_4$, which was yet again omitted due to
its perplexed form. Continuing with our calculations, the initial
and the final value of the scalar field are extracted from Eq.
(\ref{efolds}) and Eq. (\ref{indexC1}) respectively are,
\begin{equation}
\label{scalarfC1} \centering
\phi_f=\frac{|m-1|}{\kappa}\sqrt{\frac{2}{\omega}}\, ,
\end{equation}
\begin{equation}
\label{scalariC1} \centering
\phi_i=\frac{|m-1|e^{-\frac{N}{m-1}}}{\kappa}\sqrt{\frac{2}{\omega}}\,
.
\end{equation}
Choosing appropriately the free parameters of the model, it can
yield compatible results with the observations. Assuming for
example that ($\omega$, $\lambda$, $m$, $c_1$) are equal to
(1, 1, 12, 4.4512$\times 10^{-15}$) in reduced Planck units,
viable results are produced, since the values
$n_{\mathcal{S}}=0.965$ for the spectral index of primordial
curvature perturbations and $r=0.00029265$ for the
tensor-to-scalar ratio are both accepted values. In addition, the
spectral index of tensor modes is equal to $n_T=-0.00003658$ which
as expected has a very small value. Similarly, the values of the
scalar field from Eq. (\ref{scalariC1}) and Eq. (\ref{scalarfC1})
are $\phi_i=0.665317$ and $\phi_f=15.5563$ in Planck Units. In
this case, the scalar field increases as time flows, in contrast
to the models of the previous section.

Lastly, we note that all the approximations made in this power-law
model, both the slow-roll approximations and the ratio
$\xi'/\xi''$ hold true. We note that at the start of inflation the
slow-roll approximation seems valid, since  $\dot
H/H^2\sim\mathcal{O}(10^{-4})$ and the kinetic term of the scalar
field as well is insignificant compared to the scalar potential as
$\frac{1}{2}\omega\dot\phi^2\sim\mathcal{O}(10^{9})$ and
$V(\phi)\sim\mathcal{O}(10^{14})$. Finally, the condition
(\ref{maincondition}) must also be investigated if it holds true,
so by choosing $(\omega, \lambda,m,c_1)=(1, 1, 12,
4.4512\times 10^{-15})$ we have, $V'/V\sim\mathcal{O}(10^{3})$,
$\xi'V\sim\mathcal{O}(10^3)$ while
$\xi'/\xi''\sim\mathcal{O}(10^{-3})$. Thus the term $\xi'/\xi''$
is indeed insignificant compared to the other terms entering the
differential equation (\ref{maindiffeqn}).

Now let us proceed to some examples for which the viability with
the observational data cannot be achieved. Let us now examine the
case where,
\begin{equation}
\label{1} \centering \xi(\phi)V(\phi)=\Lambda\, ,
\end{equation}
This assumption simplifies again Eq. (\ref{maindiffeqnslowroll})
which now reads
\begin{equation}
\label{2} \centering V'(\phi)(1-\Lambda\frac{4}{3}\kappa^4)=0\,
,
\end{equation}
This particular differential equation can be interpreted in two
ways. Either the expression in the parenthesis is zero, meaning
that $\Lambda=\frac{3}{4\kappa^4}$, which is equivalent to the
previous case with $c_1=0$, or the derivative of the scalar
potential is equal to zero. The latter case requires that the
coupling function is also independent of $\phi$, so in this case
we are lead to physical inconsistencies, since the expressions
proportional to the ratio $\xi'/\xi''$ cannot be defined. Thus,
the only reasonable explanation is to assume the same power-law
model and demand that the integration constant is exactly zero.
Consequently,
\begin{equation}
\centering \xi(\phi)=\lambda(\kappa\phi)^m\, ,
\end{equation}
\begin{equation}
\centering V(\phi)=\frac{\Lambda}{\xi(\phi)}\, ,
\end{equation}
As a result, the equations for the slow-roll indices and the
expressions for the values of the scalar field at the start and
the end of inflation remain the same, where now $c_1=0$ in
slow-roll indices $\epsilon_5$ and $\epsilon_6$, and so we can
proceed directly with the evaluation of the observed quantities,
by designating properly the free parameters. However we must keep
in mind that now, the number of free parameters is reduced by one,
since Eq. (\ref{2}) demands that
$\Lambda\frac{4}{3}\kappa^4=1$. Unfortunately, there exists no
proper set of parameters which can lead to viability. It turns out
that compatibility may be achieved, only if the arbitrary
integration constant $c_1$ derived from Eq. (\ref{V1}) has a
non-zero value.

Let us briefly discuss another model, in which is related to the
string motivated Einstein-dilaton gravity, in which case the
coupling scalar function now is defined as,
\begin{equation}
\centering \xi(\phi)=Y e^{\alpha(\kappa\phi)}\, .
\end{equation}
In this case, we shall not derive the formula for the scalar
potential but we shall work only with the first slow-roll index
from Eq. (\ref{index1}). Subsequently, this particular index has
the following form,
\begin{equation}
\centering \epsilon_1\simeq\frac{\omega}{2\alpha^2}\, ,
\end{equation}
It turns out that $\epsilon_1$ is $\phi$ independent, therefore,
it is certain that this model leads to eternal inflation, if
$\alpha\gg 1$, or to no inflation at all if $\alpha\ll 1$, like
the canonical scalar theory case with exponential potential.
However, if $\alpha\gg 1$, it may be that the first slow-roll
index and the second one, as it can be shown, are constants, but
the slow-roll indices $\epsilon_4$, $\epsilon_5$ and $\epsilon_6$
are $\phi$-dependent. Thus, it may be possible that one may assume
that the inflationary era might end when one of these acquires
values of the order $\sim \mathcal{O}(1)$. This is a possibility,
but we shall not further pursuit this issue here.

Before ending, let us comment on an interesting issue related  to
the Swampland criteria
\cite{Vafa:2005ui,Ooguri:2006in,Palti:2020qlc,Brandenberger:2020oav,Blumenhagen:2019vgj,Wang:2019eym,Benetti:2019smr,Palti:2019pca,Cai:2018ebs,Mizuno:2019pcm,Aragam:2019khr,Brahma:2019mdd,Mukhopadhyay:2019cai,Marsh:2019lhu,Brahma:2019kch,Haque:2019prw,Heckman:2019dsj,Acharya:2018deu,Cheong:2018udx,Heckman:2018mxl,Lin:2018rnx,Park:2018fuj,Olguin-Tejo:2018pfq,Fukuda:2018haz,Wang:2018kly,Ooguri:2018wrx,Matsui:2018xwa,Obied:2018sgi,Agrawal:2018own,Murayama:2018lie,Marsh:2018kub}
in the context of Einstein-Gauss-Bonnet theory. This was developed
in Ref. \cite{Yi:2018dhl}, and as it was shown, the Swampland
criteria can hold true, if the scalar Gauss-Bonnet coupling is
chosen as,
\begin{equation}\label{swamp1}
\xi(\phi)=\frac{C}{V(\phi)}\, ,
\end{equation}
however in our case, where we take the GW170817 constraints into
account, the coupling function $\xi (\phi)$ of Eq. (\ref{swamp1})
does not satisfy the differential equation (\ref{maindiffeqn}),
unless the potential has a very specific form, which is the
following,
\begin{equation}\label{veryspecificpotential}
V(\phi)=A\sec\left[\frac{\kappa  \sqrt{\omega } \left(B
\left(4  \kappa ^4 C -3\right)+\phi
\right)}{\sqrt{\frac{4}{3} \kappa ^4 C
-1}}\right]\, ,
\end{equation}
where $A$ and $B$ are integration constants. As it can be
shown, the above potential does not yield a viable phenomenology
though. In addition, if we assume that the additional condition
(\ref{maincondition}) holds true, then it can be shown that  the
coupling function $\xi (\phi)$ of Eq. (\ref{swamp1}) can satisfy
the corresponding differential equation
(\ref{maindiffeqnslowroll}) only if $C=\frac{3}{4
\kappa ^4}$, however in this scenario too the model is not a
viable inflationary model, as we showed earlier in this section
(see Eq. (\ref{1})), since it leads to incompatible observational
indices with the observational data of Planck. Nevertheless, in
Ref. \cite{PhysicsLetters}, we shall demonstrate that the
Swampland criteria are naturally satisfied in the context of the
GW170817 Einstein-Gauss-Bonnet theory, for general choices of the
scalar coupling function $\xi (\phi)$ and of the potential
$V(\phi)$.

\section{Conclusions}

In this work we introduced a new theoretical framework for
studying Einstein-Gauss-Bonnet theories of gravity, which results
to particularly elegant and functionally simple gravitational
equations of motion, slow-roll indices and observational indices.
Particularly, by requiring the Einstein-Gauss-Bonnet theory to be
compatible with the GW170817 event, we ended up with a constraint
on the functional forms that the scalar Gauss-Bonnet coupling
function $\xi(\phi)$ and the scalar potential $V(\phi)$ must have.
By also using the slow-roll assumption for the scalar field and
the Hubble rate, we demonstrated that the gravitational equations
of motion end up to have a very simple form, and that the scalar
Gauss-Bonnet coupling function $\xi(\phi)$ and the scalar
potential $V(\phi)$ must satisfy a differential equation.
Accordingly we calculated the slow-roll indices for the
GW170817-compatible Einstein-Gauss-Bonnet theory, and we
calculated the observational indices of inflation too. With regard
to the latter, we focused on the spectral indices of scalar and
tensor perturbations and the tensor-to-scalar ratio. We applied
the formalism we derived in several models of interest, and we
confronted the models directly with the observational data coming
from the latest Planck 2018 results. Particularly, the most
interesting model has a scalar Gauss-Bonnet coupling function
$\xi(\phi)$ related to the error function. As we showed, this
model and a generalized model based on this, is compatible with
the Planck 2018 data, for a wide range of the free parameters
values. In addition, all the models we presented satisfy all the
constraints imposed by the slow-roll and additional assumptions,
made for the derivation of the gravitational equations of motion.

More interestingly, we investigated the phenomenological
implications of the additional condition $\xi'/\xi''\ll 1$, which
is motivated by the slow-roll conditions that are assumed to hold
true. As it turns, the resulting differential equation that
constrains the functional form of the scalar Gauss-Bonnet coupling
function $\xi(\phi)$ and of the scalar potential $V(\phi)$,
becomes simpler in this case, and this opened a new window for
obtaining interesting inflationary phenomenology. We presented
three models of interest, in all of which we fixed the scalar
Gauss-Bonnet coupling function $\xi(\phi)$ to be a power-law type,
exponential and of the form $\xi (\phi)\sim 1/V(\phi)$. The
power-law type of model was demonstrated to be compatible with the
observational data, while the last two were found incompatible
with the observational data. We also further discussed in brief
the case $\xi (\phi)\sim 1/V(\phi)$ which is related to the
Swampland in the context of Einstein-Gauss-Bonnet theories. As we
showed, the functional form $\xi (\phi)\sim 1/V(\phi)$ is not
compatible with the GW170817 results, unless the potential has a
very specific form, which however leads to non-viable inflationary
phenomenological results. However, as we show in another work
\cite{PhysicsLetters}, the Swampland criteria are compatible with
the GW170817-compatible Einstein-Gauss-Bonnet theories for quite
general functional forms of the scalar Gauss-Bonnet coupling
function $\xi(\phi)$ and of the scalar potential $V(\phi)$.

In principle, more elaborate potentials can also produce quite
interesting phenomenology in the context of the
GW170817-compatible Einstein-Gauss-Bonnet theories, by simply
fixing the scalar potential and seeking for the the scalar
Gauss-Bonnet coupling function $\xi(\phi)$, or vice-versa.
However, this paper was an introductory paper introducing the new
formalism, and showing that it is possible to provide
phenomenologically viable results for the inflationary era. We
hope in a future work to provide further interesting models that
yields also a phenomenologically viable inflationary era.

Another interesting question we would like to comment on before
closing is whether Einstein-Gauss-Bonnet theories can in principle
be reduced to Einstein's General Relativity in some limit. It is
possible to obtain Einstein gravity in the presence of a scalar
field during some eras in a cosmological context, if the scalar
coupling function $\xi (\phi)$ takes values in that era quite
small or nearly constant, and only if the curvature during that
era is very small. In the case of small values the
scalar-Gauss-Bonnet term can be small, and if $\xi (\phi)$ is
nearly constant, the Gauss-Bonnet term can be integrated from the
action, thus it has no effect, and we are left with Einstein
gravity in the presence of a scalar field with potential.
Nevertheless, during inflation, where the curvature is quite large
this is not possible. This smoothing to the general relativity
case can occur only during the late-time era, as was shown in Ref.
\cite{Nojiri:2005vv}, see also references therein.

However, caution is needed when considering these higher order
string motivated theories. In general, Einstein-Gauss-Bonnet
gravity is one of the two theories that violate the General
Relativity requirement that the metric tensor solely mediates
gravity \cite{Nair:2019iur}, the other theory is the four
dimensional Chern-Simons gravity, that is, a gravitational theory
with again the presence of a coupling between the scalar field and
the four dimensional Chern-Pontryagin term, which is a topological
invariant in four dimensions, but in the presence of the scalar
field coupling to it, it yields non-trivial effects, like in the
Einstein-Gauss-Bonnet case. Both the aforementioned theories are
string theory motivated, and can be the low-energy limit of a more
fundamental string theory, upon dimensional reduction of the
latter.

These theories in general, and particularly the
Einstein-Gauss-Bonnet theories, have brought along major
differences in astrophysical context. Particularly, the major
prediction is that black holes are scalarized (see for example
\cite{Nair:2019iur,Carson:2020cqb}), thus a fifth force emerges in
these theories, in addition to the violation of the strong
equivalence principle. Moreover, higher order effects in binaries
are also predicted, like scalar dipole radiation, which in effect
makes the rate of an inspiral of a binary system somewhat faster,
thus making clear the distinction between general relativity and
Einstein-Gauss-Bonnet theories \cite{Carson:2020cqb}. Thus
answering the question whether a smooth limit to Einstein gravity
can be obtained, is not easy in general.

\section*{Acknowledgments}

This work is supported by MINECO (Spain), FIS2016-76363-P, and by
project 2017 SGR247 (AGAUR, Catalonia) (S.D.O).

\subsection{Appendix:Motivation Behind the Various Choices of the Coupling Function $\xi (\phi)$}

In this Appendix we discuss in brief the motivation for the
choices of the coupling function $\xi (\phi)$. We introduce a
simple and elegant way of deriving the expression of the slow-roll
index $\epsilon_1$ and in consequence, the initial and final value
of the scalar field during inflation. Throughout our calculations,
it was shown that the coupling function appears in the form of the
ratio $\xi'/\xi''$. Hence, it is only reasonable to try and
simplify this expressions in order to facilitate our study. This
can be done easily by defining the derivative of the coupling
function $\xi(\phi)$, which mathematically speaking is assumed to
be at least three times differentiable, as,
\begin{equation}
\centering
\label{xi'}
\xi'(\phi)=\kappa\lambda e^{\int{\kappa X[\phi]d\phi}},\,
\end{equation}
where $\lambda$ is a dimensionless parameter and $X[\phi]$ is
dimensionless arbitrary expression depending on the scalar field.
This form was chosen simply because by differentiation with
respect to the scalar field, we end up with the following
expression
\begin{equation}
\centering
\xi''=\kappa X[\phi]\xi'.\,
\end{equation}
Thus, from equations (\ref{index1}) and (\ref{efolds}), we see
that the ratio which appears is replaced by the term $X[\phi]$.
Choosing appropriately this term leads to analytic expressions and
to an easily extracted phenomenology. One can choose to work with
such term in order to find an appealing and functional formula for
the initial value of the scalar field $\phi$ and then later derive
the expression of the Gauss-Bonnet coupling scalar function by
simply integrating Eq. (\ref{xi'}).

For instance, we mention that choosing
$X[\phi]=\frac{m}{\kappa\phi}$, leads to the
\begin{equation}
\centering
\xi'(\phi)=\kappa\lambda(\kappa\phi)^{m},\,
\end{equation}
which by further integration leads to the power-law form,
\begin{equation}
\centering
\xi(\phi)=\frac{\lambda}{m+1}(\kappa\phi)^{m+1},\,
\end{equation}
which was studied in the present paper. In addition, the extra
constant $m+1$ can be absorbed from $\lambda$ without altering the
resulting ratio $\xi'/\xi''$. Thus, in this formalism, one can
work differently and instead of defining the coupling function,
guess the relation between the first two derivatives of this
function and upon deriving the results, work backwards in order to
find the initial form of the coupling function responsible for
generating those results. This enables us to work with a plethora
of forms for the expressions $X[\phi]$ which would otherwise be
very difficult to produce.

\end{document}